\documentstyle[preprint,aps]{revtex}
\begin{document}
\draft
\title{Neutrino-induced nucleosynthesis and the site of the
$r$-process}
\author{W. C. Haxton$^1$, K. Langanke$^{2,3}$, Y.-Z. Qian$^3$, 
and P. Vogel$^3$}
\address{$^1$Institute for Nuclear Theory, Box 351550, 
and Department of Physics, Box 351560,\\
University of Washington, Seattle, Washington 98195\\
$^2$Institute for Physics and Astronomy, University of Aarhus,
Denmark\\
$^3$Department of Physics, 161-33, and W. K. Kellogg Radiation 
Laboratory, 106-38\\
California Institute of Technology, Pasadena, California 91125}

\date{\today}
\maketitle
\begin{abstract}
If the $r$-process occurs deep within a Type II
supernova, probably the most popular of the proposed sites,
abundances of $r$-process elements may be altered 
by the intense neutrino flux.  We point
out that the effects would be especially pronounced for 8 isotopes
that can be efficiently synthesized by the neutrino reactions 
following $r$-process freeze-out. We show that the observed abundances
of these isotopes
are entirely consistent with neutrino-induced 
nucleosynthesis, strongly
arguing for a supernova $r$-process site.  The
deduced fluences place stringent constraints on the freeze-out
radius and dynamic timescale of the $r$-process.
\end{abstract}

\pagebreak

It is known that approximately half of the heavy elements 
with mass number $A>70$ and all
of the transuranics are formed by the process of rapid neutron
capture, the $r$-process. The astrophysical site where the required
conditions occur --- neutron number
densities in excess of $\sim 10^{20}$ cm$^{-3}$ and temperatures
of $\sim 10^9$ K
lasting for on the order of 1 s \cite{mathews}
--- has been a matter of
speculation for almost four decades \cite{burbidge}.  
The suggested sites \cite{mathews}
include the
neutronized
atmospheres just above the supernova core, neutron-rich
jets from supernovae or neutron star coalescence, and an
inhomogeneous big bang. In addition to these so-called
primary sites, there are also secondary $r$-process models 
which can succeed with somewhat lower neutron number densities
and temperatures, but require
pre-existing heavy nuclei to act as seeds for the neutron capture.
Proposed secondary sites vary from the 
He and C shells during explosive burning in Type II
supernovae to
the core He flash in low-mass red giants.

In recent years a number of observational and theoretical arguments
have strengthened the case for a primary
$r$-process in Type II supernovae.  
The discovery of very metal-poor halo stars
([Fe/H] $\sim -1.7$ and $-3.12$) enriched in $r$-process elements 
with relative abundance distributions characteristic of 
the solar system argues 
that the $r$-process
is primary, already operating early in the history of the galaxy 
\cite{cowan}.
Studies of galactic chemical evolution \cite{mathews}
have found that the growth of $r$-process material is consistent with
low-mass Type II supernovae being the $r$-process site.
Finally, the suggestion made long ago that the $r$-process might be 
associated with expansion and cooling of neutron-rich matter from the
vicinity of the mass cut in supernovae \cite{hillebrandt} has been 
modeled much more convincingly. It has been shown 
in Ref. \cite{woosley} 
that an expanding neutron-rich nucleon gas can undergo an 
$\alpha$-particle freeze-out,
in which effectively all of the protons are 
consumed,
followed by an $\alpha$-process in which seed nuclei near 
$A \sim 100$ 
are produced.
The $r$-process then takes place through the capture of the excess 
neutrons on these seed nuclei.  While this specific model has some
shortcomings --- overproduction of $^{88}$Sr,
$^{89}$Y, and $^{90}$Zr and the need for very high entropies ---
it has demonstrated that a supernova ``hot bubble'' $r$-process can
produce both a reasonable elemental abundance distribution 
and an appropriate
amount of $r$-process ejecta.
  
In the model of Ref. \cite{woosley} the $r$-process freezes out at
radii of 600--1000 km and at times of $\sim 10$ s after core bounce.
It follows that the synthesis and subsequent ejection of the 
$r$-process products take place in an intense flux of neutrinos
of all flavors emitted by the cooling protoneutron star.
As it is known that neutrinos
are capable of inducing important nucleosynthesis in the 
C and O shells at much larger radii in Type II supernovae
\cite{woosley2}, clearly this neutrino fluence could
have consequences for the $r$-process. 

Neutrino reactions can affect the $r$-process in two ways,
by altering the path or pace of the nuclear flow during the
synthesis, or by modifying (postprocessing)
the abundance pattern after
freeze-out. The former possibility has been suggested and/or
discussed by several papers (see references given in
Ref. \cite{qian}), including the recent work where the effects of  
neutrinos on the charge flow were shown, under certain 
conditions, to improve the agreement with inferred
abundances \cite{gail}.
Much less work has been done on neutrino postprocessing.
In a recent, more technical paper \cite{qian} we 
re-examined many of the
neutrino physics issues affecting both the $r$-process itself
and the subsequent postprocessing.  The purpose of this 
letter is to highlight one result with broad implications,
that the site of the $r$-process might be deduced from
certain specific neutrino postprocessing signatures.

Supernova models predict that a cooling protoneutron star emits
about $3\times 10^{53}$ erg in neutrinos, with the energy 
roughly equi-partitioned among all species.  The rate of
neutrino reactions at radius $r$ from the center of the neutron star
is
\begin{equation}
\lambda_\nu \approx 
4.97\left[{L_\nu(t)\over 10^{51}\ {\rm erg\ s}^{-1}}
\right]
\left({{\rm MeV}\over\langle E_\nu\rangle}\right)
\left({100\ {\rm km}\over r}\right)^2
\left({\langle\sigma_\nu\rangle\over 10^{-41}\ {\rm cm}^2}\right)
\ {\rm s}^{-1},
\end{equation}
where $L_\nu(t)$ and $\langle E_\nu\rangle$ are 
the luminosity and average energy, respectively, 
of the neutrino species responsible for the reaction, and
$\langle\sigma_\nu\rangle$ is the corresponding cross section averaged
over the neutrino spectrum.  The neutrino luminosity is expected 
to evolve
with time as $\exp(-t/\tau_\nu)$, with $\tau_\nu \sim 3$ s.
The spectrum-averaged neutrino reaction cross
section  
sums over all final nuclear states.
The neutrino spectrum is taken to be a modified Fermi-Dirac
distribution with effective ``degeneracy parameters''
$\eta_{\nu_e}\approx\eta_{\bar\nu_e}\approx 3$ and 
$\eta_{\nu_{\mu(\tau)}}
\approx\eta_{\bar\nu_{\mu(\tau)}}\approx 0$.
The corresponding
average neutrino energies are
$\langle E_{\nu_e}\rangle\approx 11$ MeV, 
$\langle E_{\bar\nu_e}\rangle\approx 16$
MeV, and $\langle E_{\nu_{\mu(\tau)}}\rangle\approx 
\langle E_{\bar\nu_{\mu(\tau)}}\rangle\approx 25$ MeV.
  
The important reactions in Eq. (1) are the charged-current
$(\nu_e,e^-)$ reaction and the neutral-current heavy-flavor
$(\nu,\nu')$ reaction:
charged-current $\bar\nu_e$ reactions are Pauli blocked for the very
neutron-rich heavy nuclei in the $r$-process, while the lower
average energies of $\nu_e$ and $\bar\nu_e$ lead to smaller 
neutral-current cross sections.  
Our evaluation of these cross sections, described in much more detail in Ref. \cite{qian},
was based on extrapolating known nuclear responses to the 
neutron-rich nuclei of present interest, guided by explicit shell
model and continuum random phase approximation (CRPA) calculations
for certain nuclei of interest.  The $(\nu_e,e^-)$ cross
sections were treated in the allowed approximation, with the
Fermi strength $|M_F|^2 = N-Z$ carried by the isobaric analog
state and the Gamow-Teller (GT) strength $|M_{\rm GT}|^2 \sim 3(N-Z)$
carried by a broad resonance whose position and shape were determined
from nuclear systematics.  The GT strength can be
equated to the Ikeda sum rule result because the
$(\bar\nu_e,e^+)$ channel is effectively blocked by the large neutron
excesses in the nuclei of interest. Similar studies of the
charged-current $(\nu_e,e^-)$ reactions on heavy nuclei have been
carried out in Ref. \cite{fuller}. The corresponding neutral-current
calculation is more complicated as, in addition to the allowed
GT transition, forbidden transitions
become important due to the higher average heavy-flavor neutrino
energies.
The neutral-current results used in this
letter were taken from the CRPA calculations of Ref. \cite{qian}.

The charged-current and forbidden neutral-current reactions
typically produce a daughter nucleus excited well into the
continuum.  The nucleus
then de-excites by emitting one or more neutrons.
This is the process that alters the $r$-process abundance
distribution.
The average number of spallation neutrons,
$\langle n\rangle$, is obtained by folding the neutrino-induced 
excitation
spectrum with the neutron-evaporation
spectrum determined from the statistical model \cite{gavron}.
The total rates of charged-current and neutral-current
reactions on an average nucleus in the $A\sim80$, 130, and 195
regions are $\sim 9$, 15, and 20 s$^{-1}$, respectively, with the
corresponding average numbers of spallation neutrons
$\langle n\rangle\sim 2$, 2, and 3. These rates
are evaluated at the radius $r=100$ km for a luminosity of
$10^{51}$ erg s$^{-1}$ per neutrino species.
  
The $r$-process freezes out when the neutron number density
drops below a critical level.
The resulting $r$-process progenitor nuclei would, in the
absence of neutrino postprocessing, decay back to the valley of 
$\beta$-stability, producing the abundance pattern found in nature.
However, if this freeze-out occurs in an intense neutrino flux, both
charged-current and neutral-current reactions take place on 
the progenitor
nuclei (and their daughters), modifying the final $r$-process
abundance distribution in a characteristic way.  
We make three approximations 
in evaluating
these effects.  First, we exploit the fact \cite{qian} that
neutrino rates and neutron spallation yields do not vary
excessively (e.g., by more than about $\pm$ 40\%) over an abundance
peak.  (Variations between
peaks are more significant.)  Thus it is a reasonable approximation
to assign average rates and neutron emission probabilities to 
each abundance peak.  Second, we employ
these mean progenitor rates and neutron emission probabilities
throughout the postprocessing phase, even as $N-Z$
is evolving due to $\beta$-decay and 
neutrino reactions.  This is a good
assumption for neutral-current reactions, where rates are tied
to sum rules \cite{qian} that are only weakly dependent on $N-Z$,
but more dangerous for charged-current reactions, where the direct
dependence of rates on $N-Z$ could generate important corrections
if the number of $\beta$-decay or neutrino reactions is large
during postprocessing.
However, for the fluences we
consider below, the mean number of postprocessing neutrino reactions
is less than unity.
Third, we do not account for
the subsequent processing of neutrons liberated in the spallation.
Because the effects of reabsorption are spread over a broad range of 
$r$-process
nuclei, they are of minor importance to the 8 special
``window nuclei'' we discuss below.

With these approximations, the neutrino postprocessing effects for
a given abundance peak can be evaluated without reference to the 
details of the $r$-process freeze-out pattern or of the decay-back to
the valley of $\beta$-stability.
These effects depend only on the total neutrino
fluence through the $r$-process ejecta following freeze-out.  
Our results will be given in terms of the dimensionless parameter
${\cal {F}}$, the fluence in units of 10$^{47}$ erg km$^{-2}$,
and can be immediately applied to any hydrodynamic $r$-process
scenario for which the neutrino postprocessing fluence is
known.  Clearly, ${\cal{F}}$
depends on the 
radius $r_{\rm FO}$ and neutrino luminosity $L_{\nu,\rm FO}$
at freeze-out, 
and the time over which a significant neutrino
irradiation continues, which in turn depends on both 
the outflow velocity $v$ of the ejecta and the neutron star cooling 
history.  For example, in a neutrino-driven wind scenario
\cite{qian2} the outflow can be described 
by a constant dynamic timescale $\tau_{\rm dyn}=r/v$, i.e.,
$r\propto\exp(t/\tau_{\rm dyn})$.  With 
$L_\nu\propto\exp(-t/\tau_\nu)$, we have
\begin{equation}
{\cal{F}}={1\over 2}\left({L_{\nu,\rm FO}\over
10^{51}\ {\rm erg\ s}^{-1}}\right)\left({100\ {\rm km}\over
r_{\rm FO}}\right)^2\left({\tau_{\rm dyn}\over{\rm s}}\right)
{1\over 1+\tau_{\rm dyn}/(2\tau_\nu)}.
\end{equation}
  
The remaining calculations involve rather straight-forward 
combinatorics, described in more detail in Ref. \cite{qian}.  One
first determines $\bar{N}(n)$, the mean number of neutrino events 
(including both charged-current and neutral-current reactions, 
which prove to
be of comparable importance) producing exactly $n$ neutrons in the 
subsequent spallation after
freeze-out.  Each $\bar{N}(n)$ is proportional to
the fluence ${\cal{F}}$.
Under the assumptions enumerated above, 
the rates and neutron emission
probabilities in the vicinity of each abundance peak 
are not affected by the prior history of the target
nucleus.  Thus the distribution of
neutrino events that produce
exactly $n$ spallation neutrons is governed by a Poisson
distribution with parameter $\bar{N}(n)$.  The overall probability
for a given nucleus to emit, for example, two neutrons can
then be evaluated by listing the ways this can be done (e.g.,
two neutrons can be produced by one interaction that knocks out
two neutrons, or by two interactions each of which knocks out one),
and folding the Poisson probabilities for each type of events
in the product. The probability $P_n$ for an average nucleus
in the $A\sim 195$ region to emit a total of $n$ neutrons
after freeze-out is illustrated in Fig. 1 for three different
values of ${\cal{F}}$. The bumps in the probability distributions
at $n=4$ and 5 in this figure are due to the charged-current
$(\nu_e,e^-)$ reactions, which tend to knock out more neutrons after
each reaction.

The most straightforward use of these probabilities would be to
include them in a standard $r$-process network calculation. However,
there is an alternative and very instructive use of these results
that does not require a 
base-line $r$-process freeze-out pattern from theory:
begin with the $r$-process abundance distribution observed
in nature and, for a
given neutrino fluence, invert this distribution to determine
the initial distribution that must have existed prior to the neutrino
postprocessing.
This initial distribution would be the one conventional theory should
strive to match, if indeed we have picked the correct ${\cal{F}}$.
The appeal of this procedure is that the final $r$-process
abundances are rather tightly constrained by observation and
the neutrino physics is relatively simple, compared with other
aspects of the $r$-process.  Thus we can derive the unpostprocessed
distribution with some confidence.  
The inversion is easily carried out iteratively, as described
in Ref. \cite{qian}.
(Note that this procedure is valid even in the presence of 
$\beta$-delayed
neutron emission given the approximations detailed above.)

The dominant features of the observed $r$-process abundance
distribution
are the abundance peaks at $A \sim 130$ and 195, 
corresponding to the
progenitor nuclei with $N = 82$ and 126 closed neutron shells.
Independent of the exact value of the neutrino
fluence, the most important result of the inversion described
above is the discovery that 8 nuclei, lying in
the windows $A = 124$--126 and 183--187, are unusually 
sensitive to the neutrino postprocessing.  These nuclei sit
in the valleys immediately below the abundance peaks which can be
readily filled by spallation off the abundant isotopes in the
peaks.  This situation is entirely analogous to other cases 
where the neutrino-induced synthesis is known to be important 
\cite{woosley2}.

This observation allows us to place upper bounds on the fluence
${\cal{F}}$ characterizing 
the freeze-out of the abundance peaks.
This is done by requiring that the neutrino-induced
synthesis by itself not overproduce these nuclei.
For the $A\sim 130$ peak, we find
${\cal{F}} \lesssim  0.045$.
The limiting fluence would produce abundances of $^{124}$Te,
$^{125}$Te, and $^{126}$Te of 0.24, 0.45, and 0.65, respectively,
which can be compared with the corresponding ranges deduced 
in Ref. \cite{kaeppeler},
$0.215 \pm 0.020$, $0.269 \pm 0.042$, and $0.518 \pm 0.126$.
Thus all three isotopes would be overproduced, with the
discrepancy for $^{125}$Te being particularly severe (4$\sigma$).

In deriving this limit, a rather surprising observation was made:
a fluence slightly below this limiting value would produce
abundances in good agreement with observation.
To test the hypothesis that these three isotopes might be neutrino
postprocessing products, we repeated the inversion with the 
constraint of zero freeze-out abundances.
The postprocessed abundance distributions in the $A=124$--126
window are shown in Fig. 2 for ${\cal{F}}=0.020$, 0.031, and 0.045.
For the choice ${\cal{F}}=0.031$,
all three nuclei are produced within 
$\sim1\sigma$ of the observed abundances. 
Therefore, if a realistic $r$-process network
calculation gives a characteristic freeze-out pattern with
severely underabundant nuclei in the window, the observed
abundances of these nuclei would strongly favor an $r$-process
site with a neutrino fluence close to ${\cal{F}}=0.031$
after the freeze-out of the $A\sim 130$ peak.
Furthermore,
the unpostprocessed abundance distribution outside
the window derived by the inversion necessarily depends on
${\cal{F}}$. Thus, in principle, the comparison of this
distribution with the one calculated by the $r$-process theory
could provide a consistency check on whether we
have picked the correct ${\cal{F}}$.

The existence of a second set of postprocessing-sensitive nuclei,
those residing in the $A = 183$--187 valley just below the
$A \sim 195$ peak, provides an important additional test of the 
hypothesis that neutrino postprocessing has modified the 
$r$-process abundance distribution.  
This second window corresponds to
the stable nuclei $^{183}$W, $^{184}$W, $^{185}$Re, $^{186}$W,
and $^{187}$Re.  As in the case of the $A \sim 130$ peak we 
first establish a 
conservative
upper bound on the neutrino fluence, ${\cal{F}}\lesssim0.030$,
by finding the unpostprocessed
conditions under which all of these nuclei are overproduced 
by the postprocessing alone.
A fluence saturating this bound overproduces all five species,
with the deviations being $\gtrsim 3 \sigma$ in four cases
(and with the disagreement for $^{187}$Re being particularly
large, 0.067 compared with 0.0373 $\pm$ 0.0040 \cite{kaeppeler}).
Next, we again test the ansatz that these special nuclei  
might be the exclusive products of neutrino-induced synthesis.
The postprocessed abundance distributions of these nuclei
are also shown in Fig. 2 for ${\cal{F}}=0.007$, 0.015, and 0.030.
The choice ${\cal{F}}=0.015$
yields an excellent fit, again agreeing with observation within
$\sim 1\sigma$: the resulting abundances for $A=183$--187 are
0.0053, 0.0093,
0.0160, 0.0274, and 0.0411, respectively, which can be compared
with the corresponding observed values of 
$0.0067 \pm 0.0016$, $0.0135 \pm 0.0035$,
$0.0127 \pm 0.0024$, $0.0281 \pm 0.0024$, 
and $0.0373 \pm 0.0040$ \cite{kaeppeler}.

It is remarkable that the 8 isotopes we initially 
identified as having great sensitivity to neutrino postprocessing
prove to have abundances fully consistent with 
neutrino-induced synthesis
during postprocessing.  We consider this as strong evidence 
suggesting that
the $r$-process does occur in an intense neutrino fluence,
and thus that the interior region of a Type II supernova is
the site of the $r$-process.  The best-fit fluences derived, 
${\cal{F}}=0.031$ and
0.015, are typical of such sites.  For example, the 
$r$-process model in
Ref. \cite{woosley} is characterized by $L_{\nu,\rm FO}\sim
10^{51}$ erg s$^{-1}$,
$r_{\rm FO}\sim 600$--1000 km, and 
$\tau_{\rm dyn}\sim\tau_\nu\sim 3$ s,
yielding ${\cal{F}} \sim 0.01$--0.03.

If this conclusion is correct, neutrino-induced synthesis places
stringent new constraints on models of the $r$-process.  The
product of the neutrino flux and dynamic timescale at freeze-out
for each abundance peak is now determined, and would appear to 
require either fairly large freeze-out radii, as in Ref.
\cite{woosley},
or fairly short dynamic timescales, as deduced in 
Ref. \cite{qian2}. Our 
results also suggest that the $A \sim 195$ peak freezes out 
either at a
smaller neutrino luminosity corresponding to a later time,
consistent with Ref. \cite{woosley}, or at a larger radius
and a larger neutrino luminosity 
corresponding to a shorter dynamic timescale, as in the wind
scenario of Ref. \cite{qian2},
than the $A \sim 130$ peak.
The possibility of deriving strong constraints on the
dynamics of the $r$-process should provide adequate motivation 
for fully incorporating neutrino interactions into the networks
modeling the $r$-process and the subsequent decay-back to the
valley of $\beta$-stability.

We thank George Fuller, Brad Meyer, and Friedel Thielemann for
helpful discussions.  This work was supported in part by the
US Department of Energy and the National Science Foundation.
Y.-Z. Qian is supported by the D. W. Morrisroe Fellowship at
Caltech.

\pagebreak

\pagebreak

\begin{figure}
\caption{Postprocessing neutron emission probabilities
for an average nucleus in the $A\sim 195$ region. The
points connected by the long-dashed, dot-dashed, and
short-dashed lines are for ${\cal{F}}=0.015$, 0.030, and 0.060,
respectively.}
\end{figure}

\begin{figure}
\caption{Postprocessed abundance distributions in the
$A=124$--126 and 183--187 windows. The short-dashed,
long-dashed, and dot-dashed lines correspond to
${\cal{F}}=0.020$ (0.007), 0.031 (0.015), and 0.045 (0.030),
respectively, for the $A=124$--126 (183--187) window.
The observed abundances of Ref. {\protect\cite{kaeppeler}}
are plotted as filled circles
with errorbars. The unpostprocessed abundances in the
windows were set to zero (solid lines).}  
\end{figure}

\pagebreak

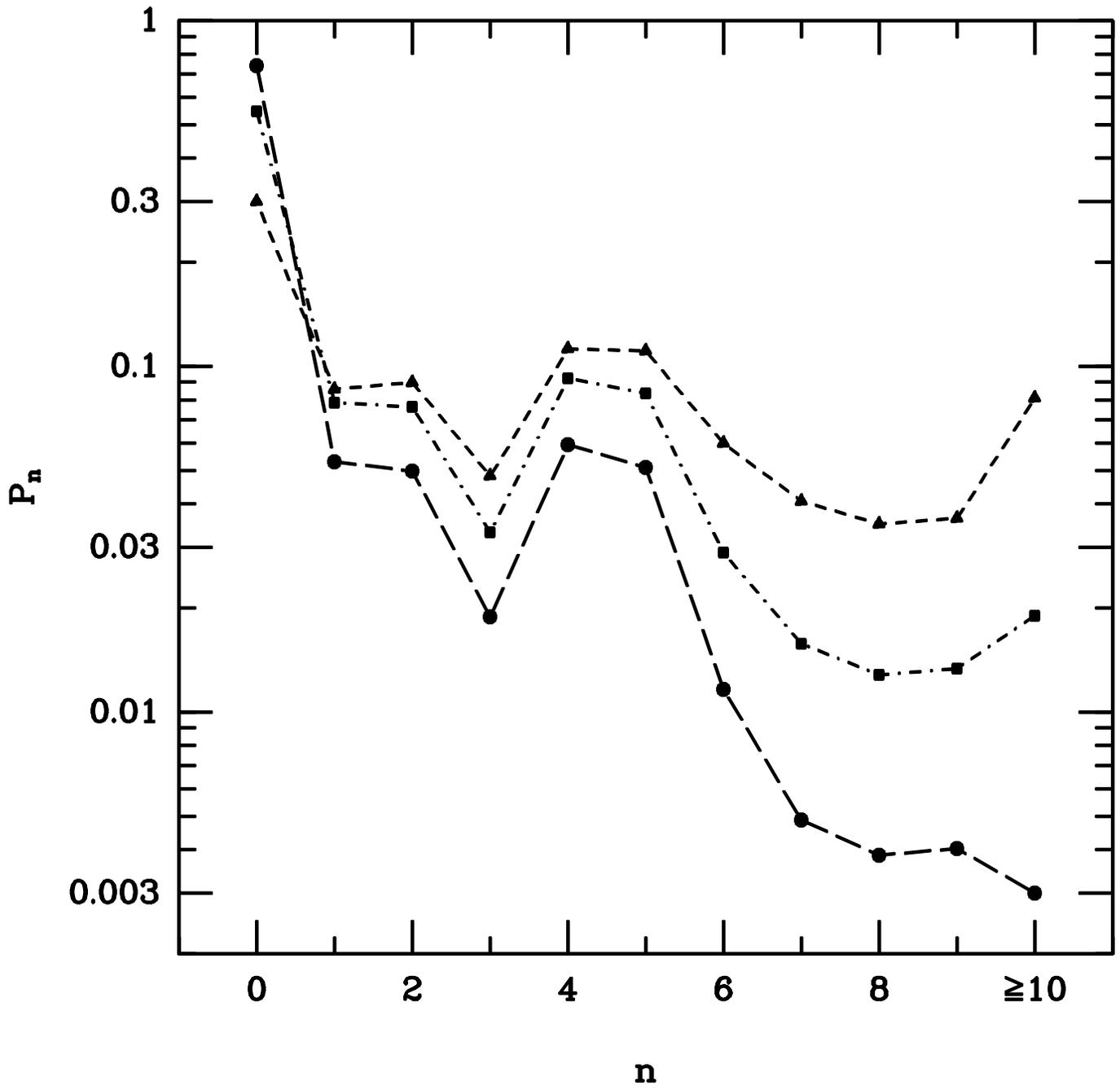
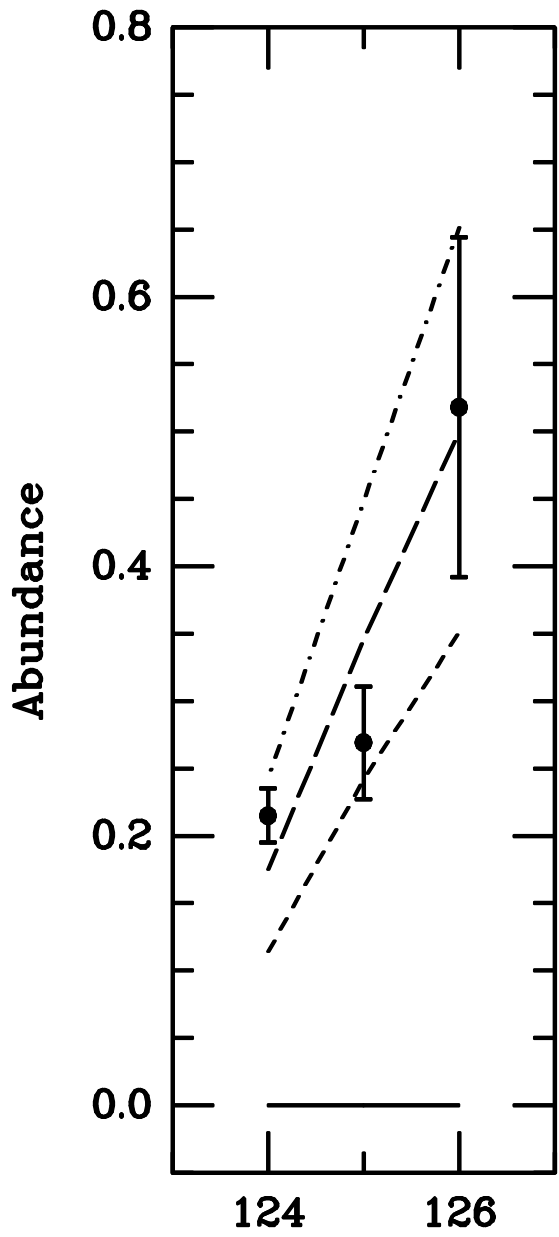
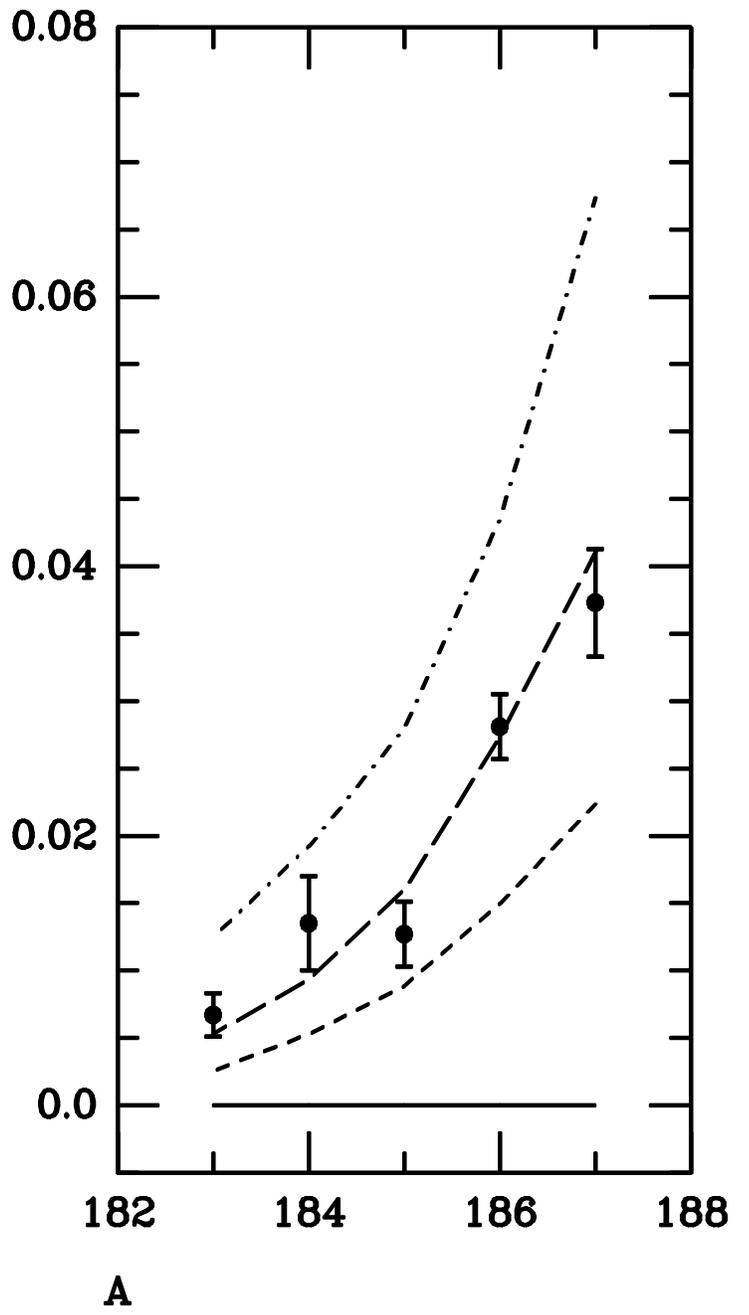
\end{document}